\documentclass{aa}
\usepackage{graphicx}
\usepackage{txfonts}
\usepackage{natbib}
\begin{document}
%
\title{Temporal variations of the CaXIX spectra in solar flares}

\author{R. Falewicz\inst{1}
          \and
           P. Rudawy\inst{1}
           \and
           M. Siarkowski\inst{2}
}
  \titlerunning{Temporal variations of the CaXIX in solar flares}

   \offprints{R. Falewicz}

   \institute{Astronomical Institute of Wroc{\l}aw University,
             51-622 Wroc{\l}aw, ul. Kopernika 11, Poland\\
             \email{falewicz@astro.uni.wroc.pl; rudawy@astro.uni.wroc.pl}
         \and
             Space Research Centre, Polish Academy of Sciences, 51-622 Wroc{\l}aw,
   ul. Kopernika 11, Poland\\
              \email{ms@cbk.pan.wroc.pl}}

   \date{Received 29 June 2009 / Accepted 18 September 2009}

\abstract{}{Standard model of solar flares comprises a bulk
expansion and rise of abruptly heated plasma (the chromospheric
evaporation). Emission from plasma ascending along loops rooted on
the visible solar disk should be often dominated, at least
temporally, by a blue-shifted emission. However, there is only a
very limited number of published observations of solar flares having
spectra in which the blue-shifted component dominates the stationary
one. In this work we compare observed X-ray spectra of three solar
flares recorded during their impulsive phases and relevant synthetic
spectra calculated using one-dimensional hydro-dynamic numerical
model of these flares. The main aim of the work was to explain why
numerous flares do not show blue-shifted spectra.}{We synthetised
time series of BCS  spectra of three solar flares in various moments
of their evolution from the beginning of the impulsive phases beyond
maxima of the X-ray emissions using 1D numerical model of the solar
flares and standard software to calculate BCS synthetic spectra of
the flaring plasma. The models of the flares were calculated using
observed energy distributions of the non-thermal electron beams
injected into the loops, initial values of the main physical
parameters of the plasma confined in the loops and geometrical
properties loops' estimated using available observational data. The
synthesized BCS spectra of the flares were compared with the
relevant observed BCS spectra.}{Taking into account the geometrical
dependences of the line-of-sight velocities of the plasma moving
along the flaring loop inclined toward the solar surface as well as
a distribution of the investigated flares over the solar disk, we
conclude that stationary component of the spectrum should be
observed almost for all flares during their early phases of
evolution. In opposite, the blue-shifted component of the spectrum
could be not detected in flares having plasma rising along the flaring
loop even with high velocity due to the geometrical dependences
only. Our simulations based on realistic heating rates of plasma by
non-thermal electrons indicate also that the upper chromosphere is
heated by non-thermal electrons a few seconds before beginning of
noticeable high-velocity bulk motions, and before this time plasma
emits stationary component of the spectrum only. After the start of
the up-flow motion, the blue-shifted component dominate temporally
the synthetic spectra of the investigated flares at their early
phases.}{}


\maketitle
%

\section{Introduction}
Solar flares are powered mainly by high-energy non-thermal electrons
accelerated somewhere in the solar corona which stream along
magnetic field lines toward the chromosphere and even photosphere
and deposit there its energy mostly in Coulomb collisions. However,
a small fraction of the energy carried by non-thermal electrons is
converted into hard X-rays (HXR) by bremsstrahlung processes
(\citet{Brown}). Rapid deposition of energy by the electron
beams cause that the energy cannot be radiated away sufficiently
fast, therefore a strong pressure imbalanced develops, and heated
plasma expands up into corona in a process known as chromospheric
evaporation \citep{Antonucci1984, fis85,Antonucci1999}.

The heated and evaporates plasma radiates over a wide
spectral range from hard X-rays or even gamma rays to radio
emission, but most of the energy is emitted in soft X-rays (SXR)
\citep{Lin1971,Petrosian1973}. Hard and soft X-ray fluxes emitted
by solar flares are generally related, in a way first described by
\citet{Neupert1968}, who found that the time derivative of the soft
X-ray flux approximately matches the microwave flux during the
flare impulsive burst. A similar effect was also observed for hard
X-ray emission (\citet{Dennis1993}). Since hard X-ray and microwave
emissions is produced by non-thermal electrons and soft X-rays are
generated by thermal emission from hot plasma, the Neupert effect
suggests that non-thermal electrons are primary source of plasma
heating.

 Detailed studies \citep{Dennis1993,Veronig2002} show
deviations from the Neupert effect, indicating that involved
processes are much more complicated. There are clear evidence of a
SXR rise before the impulsive HXR emission in some flares (often
referred to as a preheating), a persistence of the increase and
slower than expected decrease of the SXR flux is frequently observed
after the end of the HXR emission. Statistical studies (e.g.,
\cite{Lee1995, Veronig2002}) indicate additional heating
processes are necessary.

While the mechanisms and processes involved in an abrupt heating
of the flaring plasma are still not fully understood, a
potentially fruitful way to study solar flare physics is
comparison to the observational imaging and spectral data with the
results of the numerical modeling of the relevant processes.

\begin{table*}[ht!]
\caption{Physical parameters of analysed flares.} 
\label{table:1} 
\centering 
\begin{tabular}{c c c c c c c c c} 
\hline\hline 
Event     & Time of & \emph{GOES}    & \emph{GOES}         &$\gamma$ & $a_0$                & $E_c$ &  S              & $L_0$        \\ %
date      & maximum & class   & incremental  &         &                     &       &                  &              \\
          & [UT]    &         & class        &         & [ph/$cm^2$/sec/keV] & [keV] & [$10^{17} cm^2$] & [$10^{8}cm$] \\
 \hline
16-Dec-91 & 04:58   & M2.8    & M2.7         & 3.61    & $7.57\times 10^{6}$ & 25.8   & 3.85             &  15.40       \\
02-Feb-92 & 11:34   & C5.5    & C2.7         & 2.97    & $2.40\times 10^{5}$ & 18.9   & 2.30             &  09.75       \\
27-Jul-00 & 04:10   & M2.5    & M2.4         & 3.20    & $4.79\times 10^{5}$ & 19.8   & 2.31             &  08.69       \\

\hline 
\multicolumn{9}{l}{{\tiny $\gamma$ - photon spectral index; $E_c$ - low energy cut-off; $a_0$ - scaling factor (flux at 1 keV);}}\\
\multicolumn{9}{l}{{\tiny $S$ and $L_0$ - cross-section and semi-length of the flaring loop}} \\
\vspace{0.01cm}
\end{tabular}
\end{table*}

Spectra of hot and rapidly rising
plasma emission are temporally dominated by a blue-shifted
component (i.e. blue-shifted emission of the plasma dominates or
is comparable to the stationary one). It seems the strongly
blue-shifted emission should be routinely detected during flares,
at least during big ones. Most of papers describing BCS analysis
results indicate that the BCS spectra are dominated by the
emission of the stationary component. There is only a very limited
number of described observations of the solar flares having
spectra in which the blue-shifted component dominates the
stationary one (e.g., \citet{Gan02} and references therein;
\citet{Culhane94}). \citet{Gan97} performed a statistical study of
the flares having a single loop structure or a single loop
dominate the SXT images. They found that although the blue
asymmetry is very common during the impulsive phase, the number of
events with great blue-shift is very small. They also found that
for most flares the blue-shift emission appears during early
impulsive phases and is temporally correlated with the line
broadening. There exist also interesting observations of a
blue-shifted emission detected during the gradual phases of the
flares, out of the scope of this work (\citet{Czaykowska99}).

\begin{figure*}[ht!]
\centering

\includegraphics[width=5.0cm]{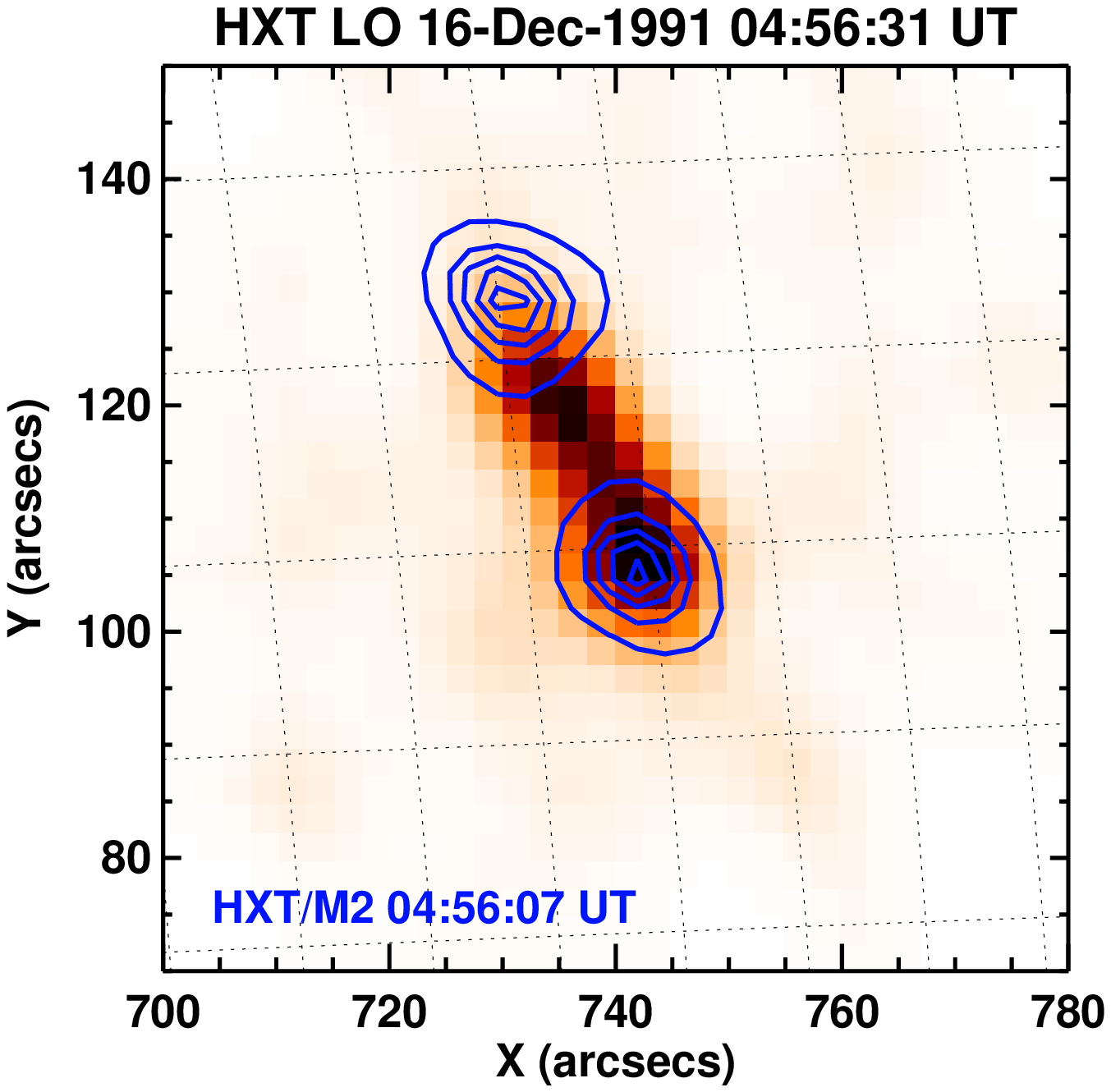}
\includegraphics[width=5.0cm]{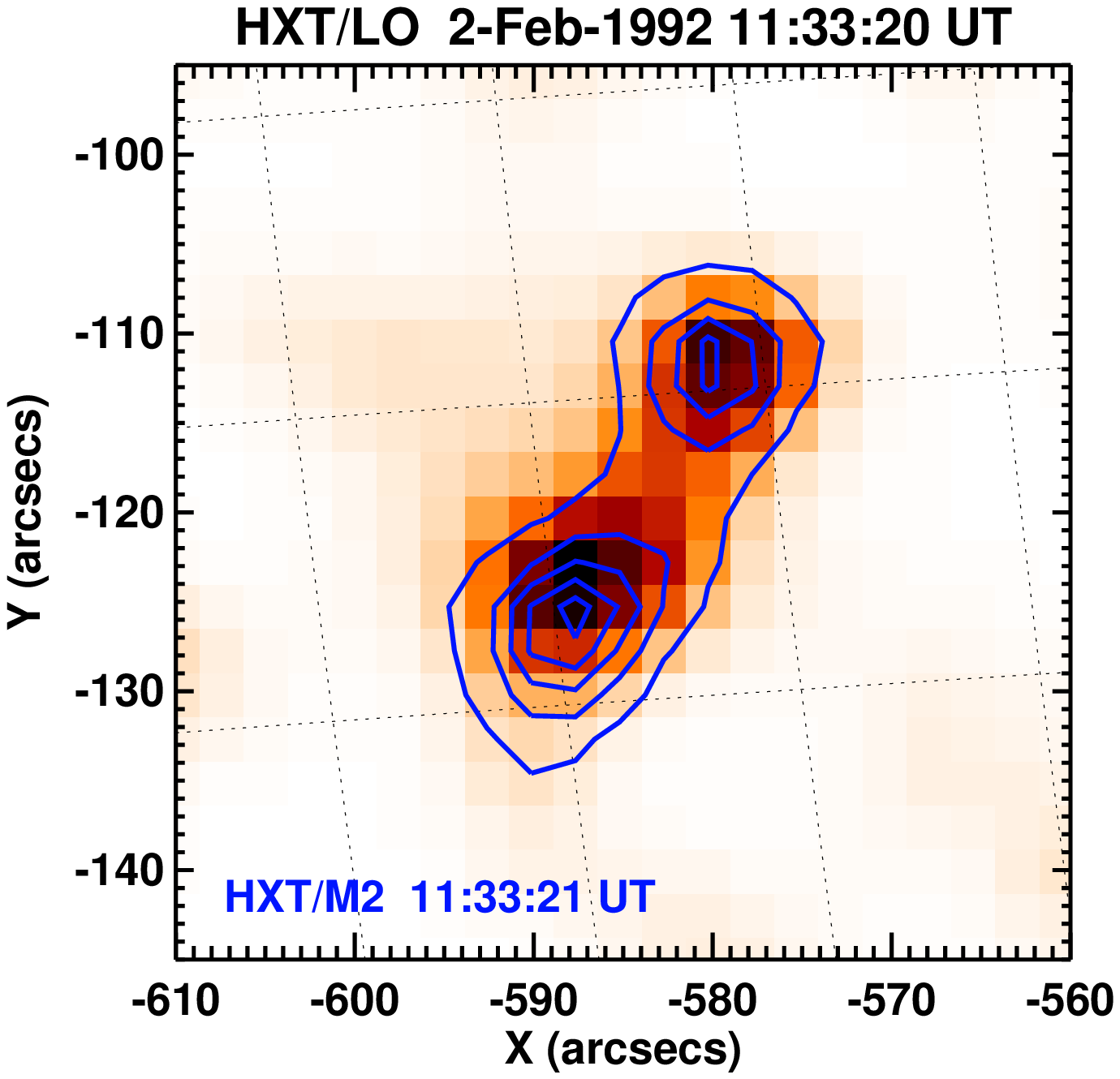}
\includegraphics[width=5.0cm]{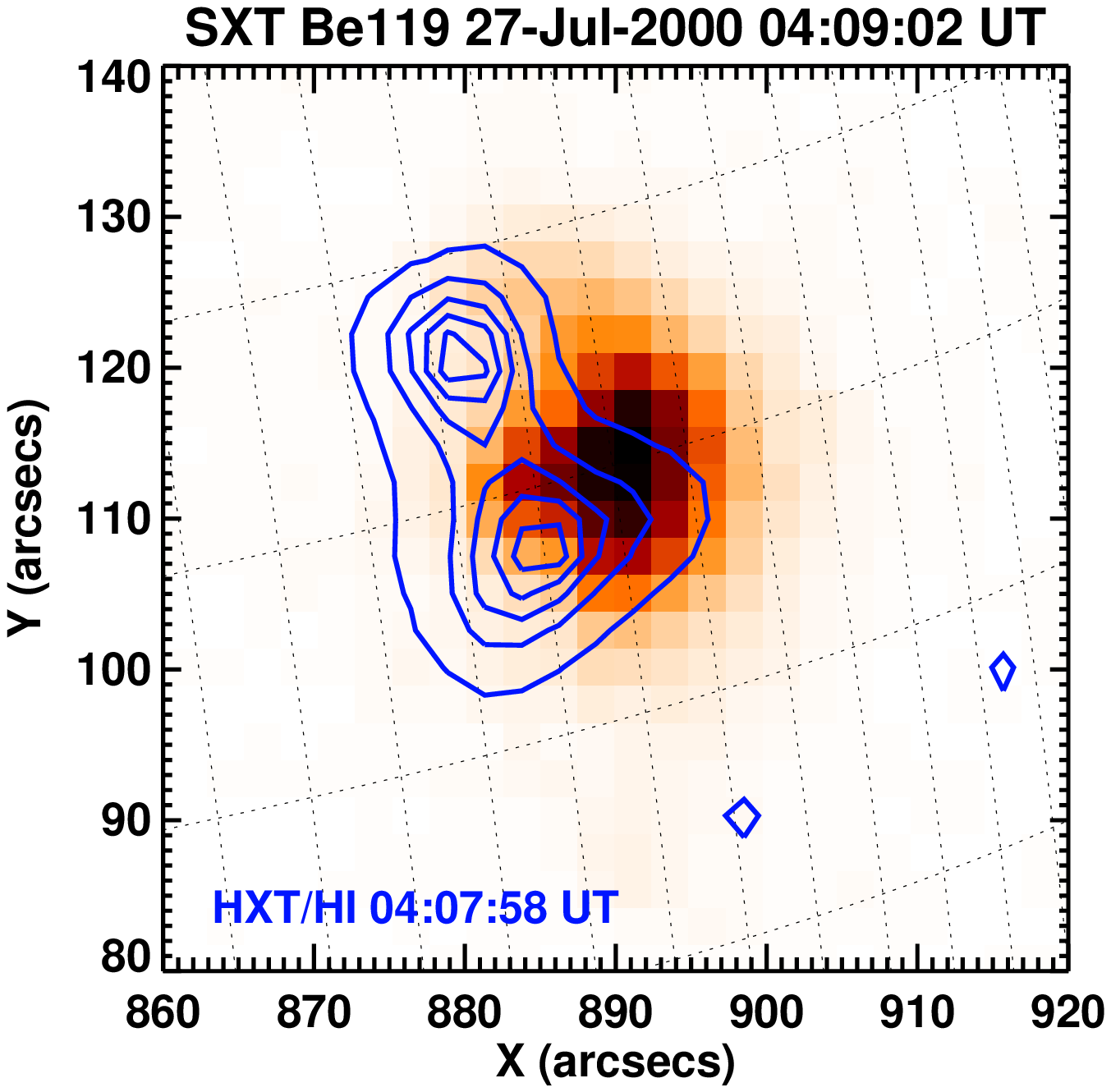}

\caption{Images of three analysed flares taken by {\it Yohkoh}
satellite with SXT or HXT/LO (gray scale images) and HXT
(iso-contours) telescopes.} \label{fig01}
\end{figure*}

Most of the published observations of the solar events did
not indicate a presence of a large blue-shifted component of the
spectrum. However,  most 1D hydrodynamic (HD) numerical models of
the chromospheric evaporation in a single-loops show a presence of
very high up-flow velocities, up to several hundred kilometers per
second \citep{Emslie1992, Mariska1991, Mariska1993, Mariska1994}.
This discrepancy between observations and models could be explained
by insufficient sensitivity of the instruments \citep{Antonucci1987,
Doschek2005} but most of the theoretical works present models of the
solar flares having low up-flow velocities. The most advanced model
of this kind is a multi-thread model of the flare proposed by
\citet{Hori1998} and developed by Doschek and Warren, consistent
with the observed BCS spectra \citep{Doschek2005, Warren2005}.
In this paper we present other explanation of the lack of
the blue-shifted spectra. For three solar flares we show that
one can well explain the observed spectra using simple 1D single-loop
HD models and taking into account geometrical effects of the viewing
angles to the loops.

We describe the observed flares (Sect. 2), the model calculation (Sect.
3), the obtained results (Sect. 4) and the discussion and
conclusions (in Sect. 5.)

\section{Observational data}

We selected three solar flares observed by the \emph{Yohkoh}
satellite on the disk, having a clearly recognizable single-loop
geometric structure and maximum hard X-ray flux not less than 10
ctns/sec per subcollimator in the M2 channel (33-53 keV) of the HXT
instrument (\citet{Kos91}). Two analysed flares: C2.7 \emph{GOES}
class flare observed on 1992 February 2 and M2.4 flare on 2000 July
27 were not mentioned previously in literature; the third one, M2.7
class flare on 1991 December 16, was investigated previously by
Culhane et al. (1994). All the flares were observed with HXT
telescope during  their impulsive phases but only one of them (2000
July 27) was also observed with the \emph{Yohkoh} SXT
grazing-incidence telescope (Tsuneta et al.  (1991)). HXT images of
the flares were reconstructed using the standard Pixon method
(\citet{Met96}) with variable accumulation times and an assumed
threshold count rate of 200 counts in the M2 band (33-53 keV). The
emissions of the flares were routinely recorded with the \emph{GOES}
X-ray photometers (1-8~ \AA~ and 0.5-4~ \AA~ bands). The images of
the investigated flares are shown in Fig. 1, and their main
characteristics are given in Table 1.

Geometrical parameters of the loops i.e. their lengths
($L_0$), cross-sections (S) and inclinations were evaluated under
assumption that an observational error of the position of the
observed structure is of the order of one pixel (i.e. 2.45 arcsec).
In a course of the calculations both semi-lengths and cross-sections
of the loops were refined (in a range of the error only) in order to
obtain the best conformity between theoretical and observed GOES and
BCS light curves (Table 1 presents final values of S and $L_0$ used
in calculations). The errors of inclinations of the loops depends on
flare localization on the Sun but for all investigated events are less than
$\pm 5$ deg.

The observed X-ray spectra of the flares were recorded with the
Bragg Crystal Spectrometer (BCS) instrument on board \emph{Yohkoh}
satellite (\citet{Culhane91}). The BCS was a full-Sun crystal
spectrometer, which measured spectra in the vicinity of strong
resonance lines of H-like FeXXVI and He-like FeXXV, CaXIX, and SXV
ions using curved crystals. The spectra were registered in
one-dimensional position-sensitive proportional counters, with a
time resolution of $<1$ sec. In this paper we used spectra measured
around CaXIX resonance line. The BCS does not have the
absolute wavelength calibration because locations of spectra on the
detector depends on the locations of the flares on the Sun. To take account
this pointing offset we determined the rest wavelengths of the
spectra using spectra recorded late in the flare when expected
velocities are small.

\subsection{M2.7 flare on 1991 December 16}

The M2.7 \textit{GOES} incremental class (i.e.,background
subtraction) solar flare occurred at 04:54 UT on 1991 December 16 in
an active region NOAA 6961 on N04W45. The flare appeared as a single
loop of moderate semi-length of about 15 400 km in SXR (see Fig. 1).
The impulsive HXR emission of the flare above 23 keV was recorded
between 04:55:50 UT and 04:56:28 UT.  We estimated the cross-section
of the loop as equal to $S=3.85\times 10^{17}$ $\rm cm^{2}$ using
the reconstructed HXR images.

The BCS observations of this flare were described in detail by
\citet{Culhane94}. We used spectra measured around CaXIX resonance
line. The spectra were recorded in a 6 sec cadence from 04:56:06 UT
except the first spectrum recorded at 04:55:42 UT with a 24 sec
integration time. A blue-shifted, highly asymmetric component
dominated between 04:56:06 and 04:56:48 UT. High asymmetry suggests
wide spectrum of up-flow velocities.

A numerical model of the flare was calculated using parameters of
the photon spectra $\gamma$ and scaling factors $a_{0}$ calculated
as a function of time on the basis of hard X-ray fluxes observed in
$M1$ and $M2$ channels of the HXT instrument.

\subsection{C2.7 flare on 1992 February 02}

The C2.7 \textit{GOES} incremental class flare on 1992 February 02
was observed between 11:33 UT and 11:38 UT in an active region NOAA
7042 on S11E41 with \textit{Yohkoh}/HXT and \textit{GOES} only (see
Fig. 1). The impulsive HXR emission above 23 keV was recorded
between 11:33:18 UT and 11:33:28 UT only. HXT/LO images of the flare
show X-ray emission of a single loop of moderate semi-length $\sim 9
750$ km. The feet of the loop were seen in images taken in HXT
channels M2 and H only. An estimated area of the loops cross-section
was equal to $S=2.30\times 10^{17}$ $\rm cm^{2}$.

The CaXIX spectra were recorded from 11:33:01 UT with 6 sec cadence.
The blue-shifted component with velocity higher than 200 km/s become
visible 6 sec later.

The numerical model of the flare was calculated in the same way as
for the previous one.

\subsection{M2.4 flare on 2000 July 27}

The M2.4 \textit{GOES} incremental class solar flare on 2000 July 27
was observed between 04:08 UT and 04:13 UT in an active region NOAA
9090 on N10W72. The flare was visible in SXT as a single loop of
moderate semi-length of about 8 690 km (see Fig. 1). The impulsive
hard X-ray emission above 23 keV was recorded between 04:07:54 UT
and 04:08:30 UT .  We estimated the cross-section of the loop as
equal to $S=2.31\times 10^{17}$ $\rm cm^{2}$ using the reconstructed
HXR images.

 The spectra were recorded with a 6 sec cadence
starting from 04:08:11 UT except the first spectrum recorded at
04:07:47 UT with a 24 sec integration time. From beginning of the
impulsive phase spectra reveal clear blue-shifted component. The
blue-shifted component was highly asymmetric and dominated between
04:56:06 UT and 04:56:48 UT. This suggests wide spectrum of up-flow
velocities.

The numerical model of the flare was calculated in the same way as
for the previous events.

\section{Method of analysis}

The BCS CaXIX synthetic spectra of the investigated events were
calculated using results of the 1D numerical HD models of the
observed events. The HD models take into account main factors
forming an energy budget and plasma kinetics of the flares: observed
energy distributions of the non-thermal electrons and temporal
variations of the emitted X-ray fluxes, geometry of the flaring
loops estimated using images of the events, estimated main initial
physical parameters of the plasma (density, temperature) and
processes of gain and losses of the energy by the matter.

\begin{figure*}[ht!]
\centering

\includegraphics[width=14.0cm]{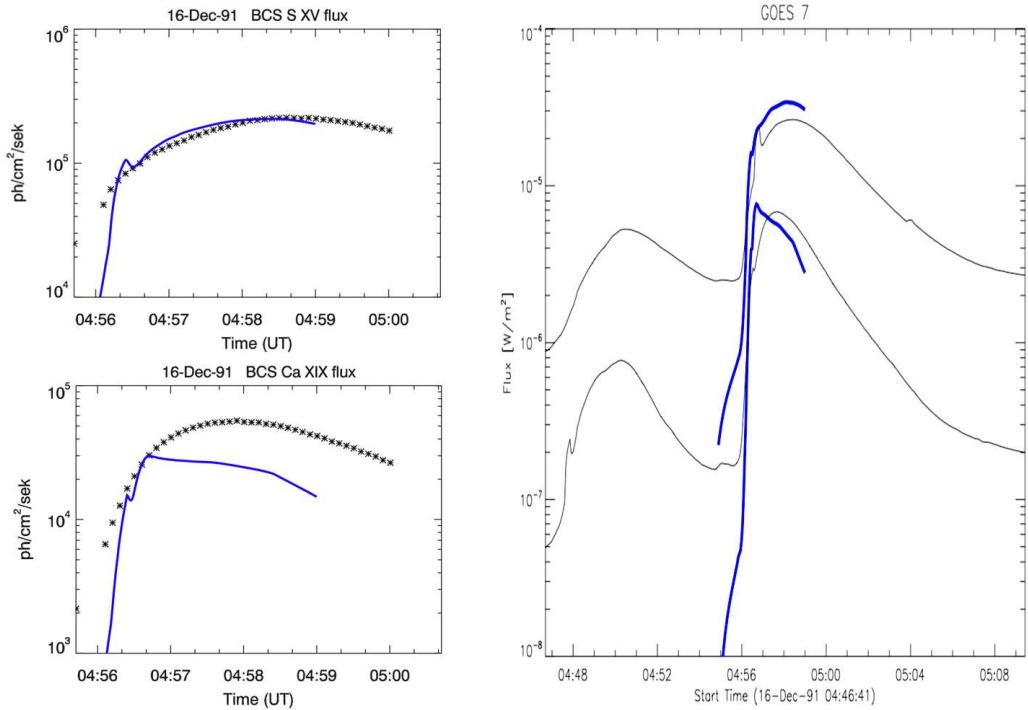}

\caption{On the left: synthesised BCS CaXIX and BCS SXV fluxes (blue
lines), and the observed ones (asterisks) of the M2.7 class solar
flare observed on 16 December 1991. On the right: synthetic X-ray
fluxes in 0.5-4~\AA~ and 1-8 \AA~ bands calculated using a numerical
model for the same solar flare observed on 16 December 1991 (blue
thick line)  and relevant fluxes recorded with \textit{GOES} 7
satellite (thin gray lines).} \label{fig02}
\end{figure*}

In this work we used the modified Naval Research Laboratory Solar
Flux Tube Model code kindly made available to the solar community by
Mariska and his co-workers (\citep{mar82, mar89}). A typical flaring
loop can be modeled for many purposes with a simple 1D hydrodynamic
model (as with the NRL Code) although it is a 3D structure
surrounded by a complex active region. We included a few modifications to the original NRL code: new radiative
loss and heating functions; the VAL-C model (\citet{Vernaz}) of the initial
structure of the lower part of the loop (extended down using Solar Standard Model data;
\citet{bahcall}), double precision of the calculations and a mesh of new values of
the radiative loss function, calculated using the CHIANTI (version 5.2) software
\citep{Dere, Landi} for a temperature range $10^4 - 10^8$ K and density range $10^8 - 10^{14} \rm cm^3$.
All changes were tested by comparison of the obtained results with the results
of original models published by Mariska.

A sole important problem which we meet during the modeling of the flares using original
NRL code was an insufficient amount of the mater located in feet of the loops.
All our attempts to apply energy flux derived from observational data of medium or
big-class solar flares caused massive evaporation of the whole NRL isothermal "chromosphere"
and an un-physical "diving" of the feet, even if the chromosphere was un-realistically thick.
To solve this problem we applied the VAL C model of the solar plasma,  extended down using
Solar Standard Model data. It was done solely in order to obtain big enough storage of the matter.
Under such assumption doesn't matter much which particular VAL model is used. All other aspects of
the NRL model of the chromosphere are unchanged (radiation is suppressed, optically thick emission
is not accounted for, and no account is taken of neutrals).

Main geometrical parameters of the flaring loops: the volume (V),
loop cross section (S), half-length ($L_0$) and a local
inclination of the loop's axis to the line of sight were
determined using images taken with SXT and HXT. The loop cross
sections were estimated as the areas within a flux level equal to
30\% of the maximum flux in the HXT/M2 channel. Loop half lengths
$L_0$ were estimated using distances between the centres of
gravity of the HXT/M2 footpoints, assuming a semi-circular shape
for the loop an fixed cross-section.

\begin{figure*}[ht!]
\centering

\includegraphics[width=12.0cm]{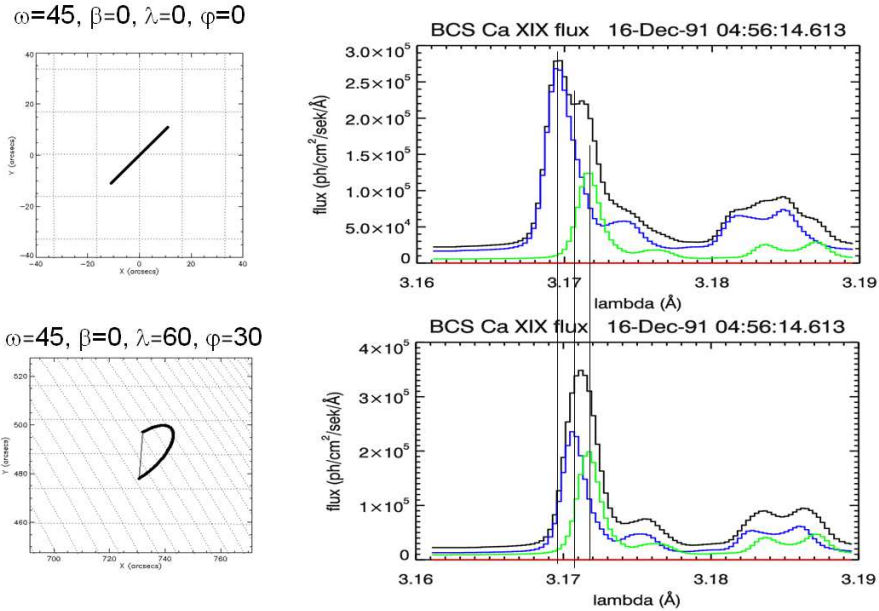}

\caption{Influence of the position and orientation of the loop on
the spectrum obtained and the observed blue-shift for the same
physical and geometrical parameters of the loop. Numerically modeled
BCS CaXIX spectra are plotted with solid thin lines (black),
blue-shifted component of the modeled spectra are plotted with blue
lines, red-shifted component with red lines and static component
with green lines, respectively.} \label{fig03}
\end{figure*}

Temperatures ($T_e$) and emission measures (EM) of the plasma were
estimated using \emph{GOES} 1 - 8~\AA~ and 0.5 - 4 ~\AA~ fluxes and
the filter-ratio method proposed by Thomas et al. (1985) and updated
by White et al. (2005). Mean electron densities (\rm $n_e$) were
estimated from emission measures (EM) and volumes ($V = 2L_{0}S$).

The heating of the plasma by the Coulomb collisions of the
non-thermal electrons was modeled using the approximation given by
Fisher (1989), necessary parameters of the non-thermal electron
distributions were calculated as a function of time from hard X-ray
fluxes observed in the HXT M1 and M2 channels. We calculated time
variations of the spectral indexes ($\gamma$) and scaling factors
$a_0$ (photon flux at 1 keV) for the impulsive phases of the
investigated flares assuming power-law hard X-ray photon spectra.
Electron spectra of the form of $F= A\,E^{-\delta}$ can be
calculated from power-law photon spectra using the thick target
approximation. A detailed description of this numerical model is
also given by \citet{Falewicz}.

All models were calculated for periods lasting from the beginning of
the impulsive phase beyond the soft X-ray emissions maximum. The
time steps in the models were about 0.0005-0.001 sec.

While the chromospheric evaporation is powered by non-thermal
electrons, a proper estimation of the energy flux carried by
electrons is decisive for realistic evaluation of the plasma
velocity field. The total energy carried by the non-thermal
electrons is very sensitive to the assumed low energy cut-off of the
electron spectrum ($E_c$), due to the power-law nature of the energy
distribution. A change of the $E_c$ value by just a few keV can add
or remove a substantial amount of energy from/to the modelled
system, so $E_c$ must be selected with great care. For investigated
flares we estimated $E_c$ using an iterative method, selecting $E_c$
value which gives the best agreement of the synthesized and observed
X-ray fluxes recorded with \emph{GOES} X-ray photometers (i.e.
\emph{GOES} classes for 1-8 ~\AA~ band) and \emph{Yohkoh}/BCS
fluxes. Examples of the achieved conformity of the observed and
synthesized BCS light curves and the observed and synthesized
\emph{GOES} curves are shown in Fig. 2. For each flare the selected
value of the $E_c$ was fixed while electron spectral index
($\delta$) and corresponding scaling factor (A) varied in time in
accordance with the observed variations of the HXR energy flux. The
values of the $\delta$ and estimated $E_c$, observed at maximum of
the HXR emission, are given in Table 1.

As a result of HD numerical modeling we obtain  time sequences of 1D
models of flaring loops. Each model consists of ~2000 cells with
individual $\rm T_e$, EM and velocity \textbf{v}. For each cell we
calculated corresponding spectrum at vicinity of CaXIX resonance
line using Chianti 5.2 emissivities for continuum and lines; each
cell's spectrum was Doppler-shifted in wavelength ($\lambda$) by
$\Delta\lambda=\lambda\cdot{v}/{c}$ according to the local
line-of-sight velocity component ($v$). The total CaXIX spectrum
from the loop at given moment of evolution was calculated as a sum
of those Doppler-shifted cell's spectra. For better comparison with
observations we used the same dispersion i.e., the wavelength bin
width as obtained from BCS data.

\begin{figure}[ht!]
\centering

\includegraphics[width=7.0cm]{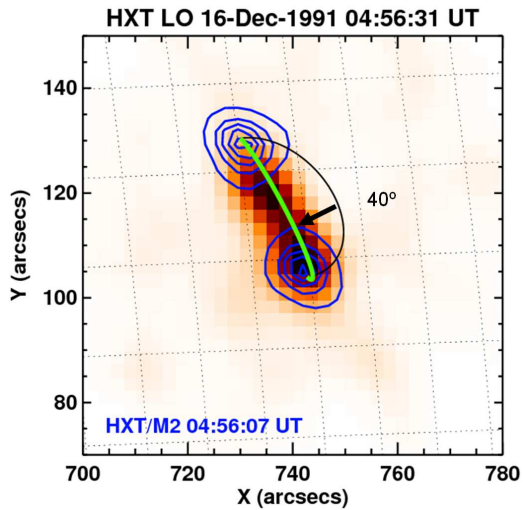}

\caption{Image of the flare taken by {\it Yohkoh} satellite with SXT
or HXT/LO (gray scale images) and HXT (iso-contours) telescopes. The
theoretical semi-circular loop anchored in the footpoints and
perpendicular to the solar surface (black) should be declined
(tilted) 40 degrees from the vertical to cover with the observed
loop (green).} \label{fig04}
\end{figure}

Velocities projected along the line-of-sight (LOS) for each volume
element (cell) of the loop were calculated in the way similar to the
method described by \citet{Li}. The LOS velocities depend on
geometry of the loop, on their orientation (inclination to the solar
surface ($\beta$) and the tilt angle ($\omega$)), and their
heliographical coordinates ($\lambda,\phi$). In Fig. 3 we compared
spectra obtained for the same velocity distribution along the loop
but for two different loop positions on the solar disk. Depending on
the loop location and orientation the blue-shifted component may be
dominant or only small feature in the spectrum.

The inclinations of the loops to the solar surface ($\beta$) were
calculated under assumptions of their semi-circular shape. Using SXT
or HXT/LO images we fitted the observed loops with semi-circular
loop having the same length and rooted in the same points, but
inclined to the solar surface (an example is shown in Fig. 4). Using
known longitudes and latitudes of the feet and inclination of the
loop to the solar surface an actual projection of a velocity vector
along the loop on the line of sight axis were evaluated for all
points of the calculation mesh along the loop.

\begin{figure*}[ht!]
\centering

\includegraphics[width=12.0cm]{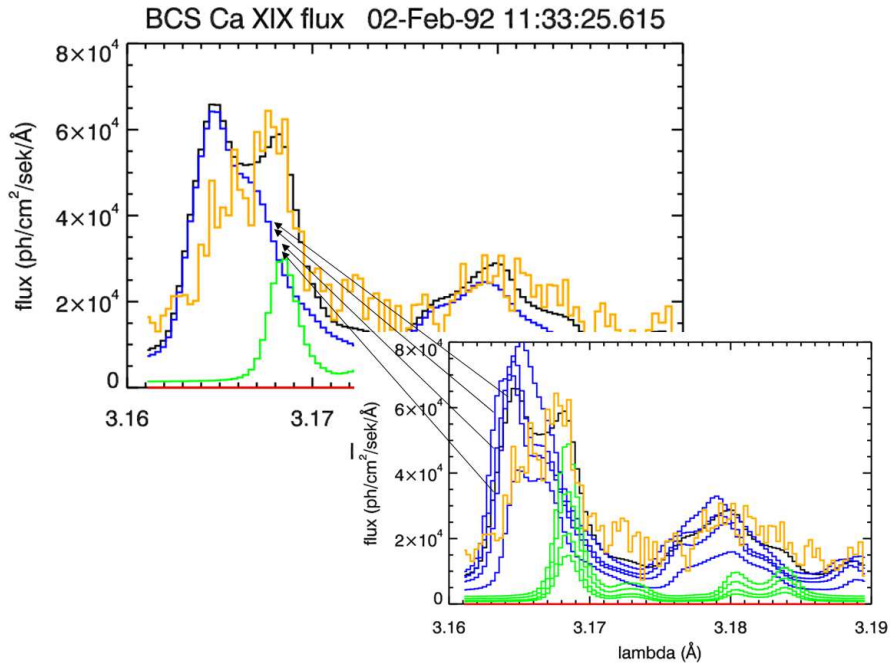}

\caption{The observed spectrum as the sum of several spectra from
the model. More details can be found in the text.} \label{fig05}
\end{figure*}

All investigated flares emitted a slightly enhanced BCS emission
long before beginnings of the detected hard X-ray emissions. In
order to generate a similarly enhanced BCS emissions in our
numerical models of the flares we started calculations of the
temporal evolution of the models tens of seconds before the
beginnings of the recorded hard X-ray emissions (i.e., before
beginnings of plasma heating by non-thermal electrons) applying a
low energy pre-heatings fitted to mimic the observed enhancement of
the BCS emissions. The values of the flux maxima are
reproduced very well but the noticeably large discrepancies during
the early rise and especially late decay phases are caused by a lack
of additional pre- and post-impulsive heating of the plasma in our model.

The fast time variations of the energy fluxes carried by non-thermal
electrons (of a sub-second time scales) cause significant changes of
the temperature, density and plasma velocity distributions along the
flaring loops and relevant variations of the HXR emission of the
solar flares in a sub-second time scale also (e.g.,
\citet{Radziszewski}). Due to the technical features of the
\emph{Yohkoh}/BCS spectrometer and limited intensities of the X-ray
emission of the investigated flares an effective time resolution of
the recorded BCS spectra was of the order of 4-6 seconds. Thus, each
recorder BCS spectrum could be recognized as a sum of numerous
spectra emitted consecutively by the flaring loop in a course of the
period of the spectrum's accumulation. In order to mimic the process
of observed spectra accumulation, the momentary BCS spectra
calculated for consecutive time steps of the numerical models of the
flares were averaged over the periods equal to the relevant periods
of accumulation of the observed BCS spectra. In Fig. 5 we present an
example of evolution of the synthetic spectra during accumulation
period. These variations are large, both stationary and blue-shifted
components can change significantly during the period when BCS
collected the spectra. This fact should be kept in mind when
interpreting the observed spectra. This is especially important
during rise phase of the flares when, because of low statistic, an
accumulation times varied from 6 up to 9 seconds.

\begin{figure*}[ht!]
\centering

\includegraphics[width=14.0cm]{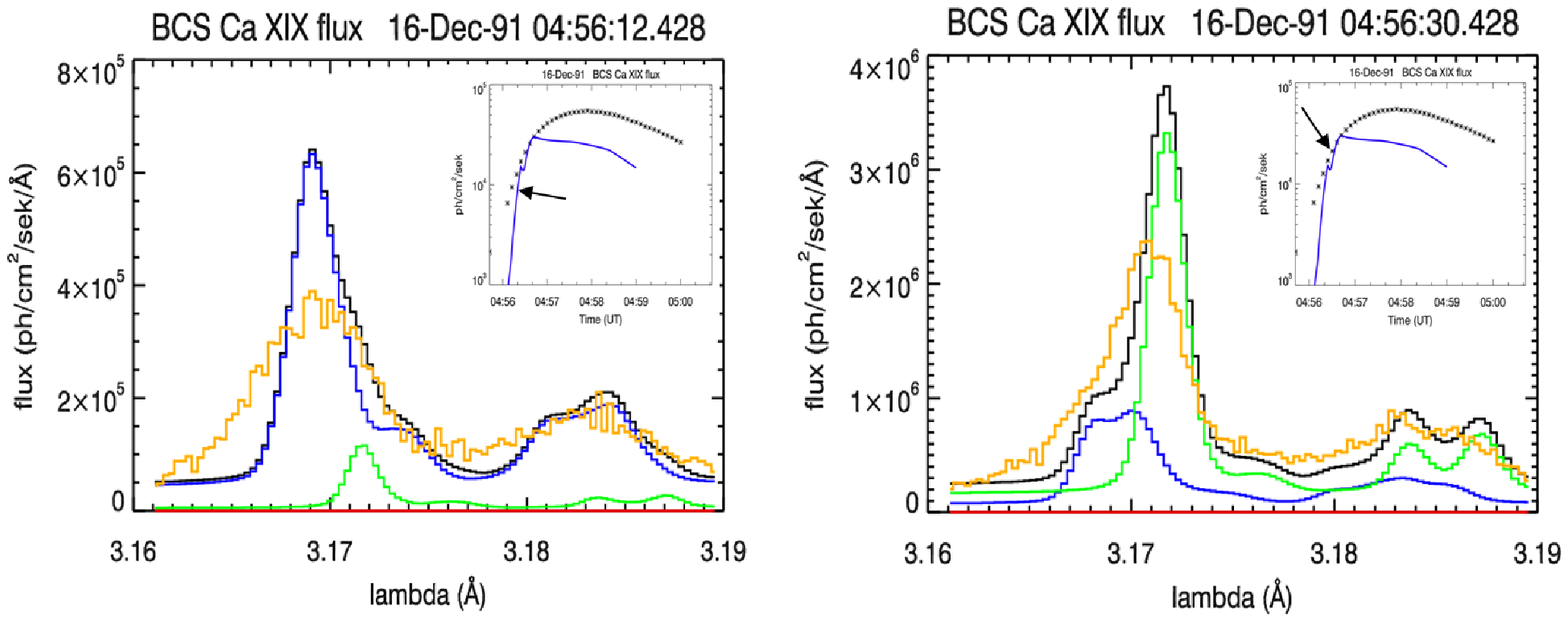}

\includegraphics[width=14.0cm]{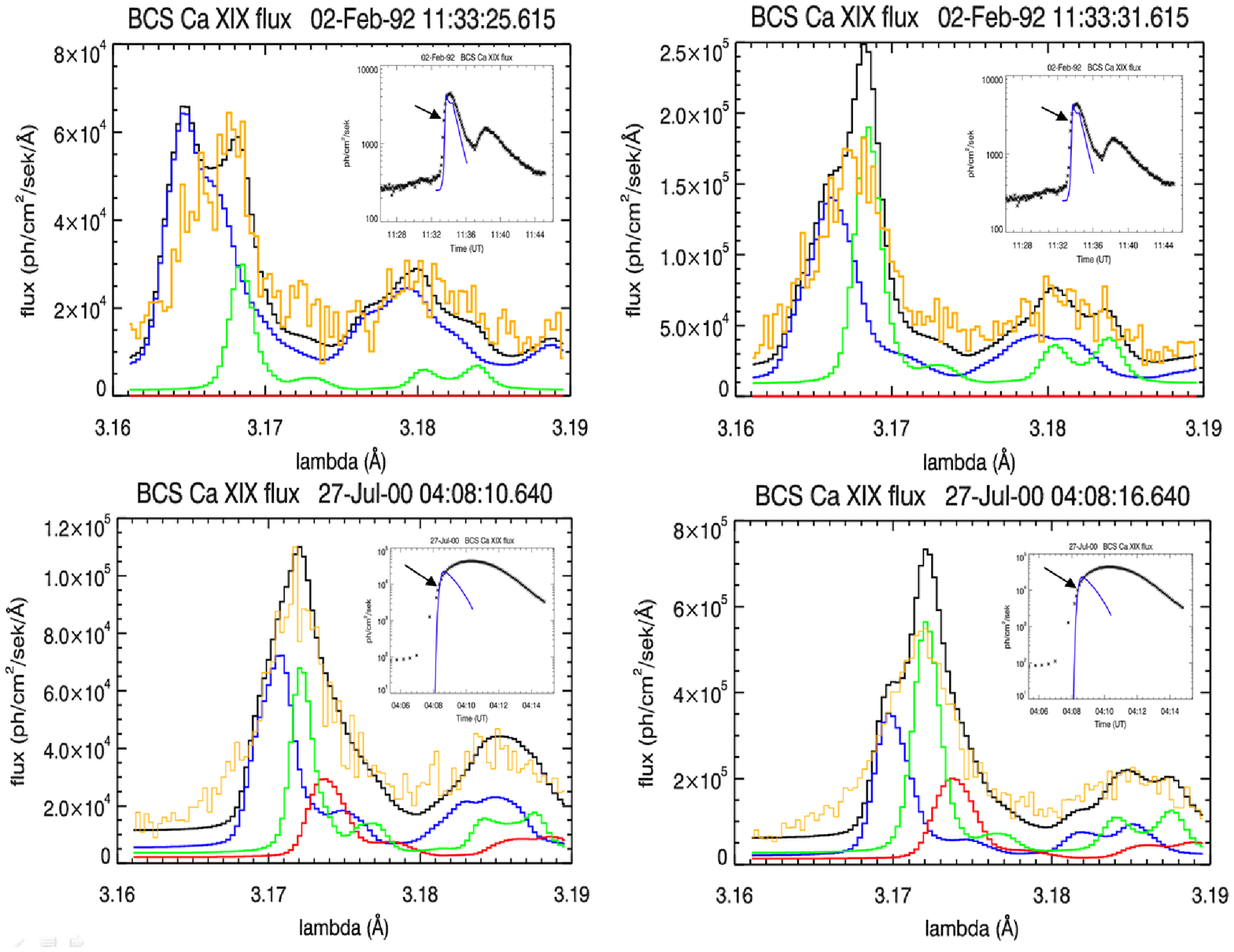}

\caption{The observed spectrum for the different moments of time for
the three analysed events. Observed BCS CaXIX spectra are plotted
using solid thick lines (yellow), numerically modeled spectra are
plotted with solid thin lines (black), blue-shifted component of the
modeled spectra are plotted with blue lines, red-shifted component
with red lines and static component with green lines, respectively.
} \label{fig06}
\end{figure*}

\section{Results}

We defined blue-shifted, static and red-shifted components of the
numerically modeled spectrum as emissions of plasma having radial
velocities: greater than 50 km/s (toward the observer), between -50
km/s and 50 km/s, and greater than -50 km/s (outward the observer),
respectively. The observed BCS CaXIX spectra are plotted using solid
thick (yellow) lines, numerically modeled spectra are plotted with
solid thin lines (black), blue-shifted component of the modeled
spectra are plotted with blue lines, red-shifted component with red
lines and static component with green lines, respectively (see Fig.
6).

\subsection{M2.7 flare on 1991 December 16}

The flare appeared in SXR as a single loop with semi-length 15 400
km inclined to the local vertical direction by 40 deg (see Fig. 1,
left panel and Fig. 4). BCS observations revealed the strong and
highly asymmetric blue-shift component recorded from beginning of
impulsive phase. High asymmetry of the spectral lines indicated wide
spectrum of up-flow velocities.

The numerical model of the flare was calculated for the period
lasting from 04:54:52 UT to 04:58:52 UT (240 sec), which covers an
impulsive phase of the flare and spreads beyond maximum of the
recorded \emph{GOES} satellite emission.

The blue-shifted emission is less than or comparable to the
stationary component only on the first observed spectrum. The time
evolutions of the synthetic and observed BCS CaXIX spectra are very
similar. The synthetic spectrum consists of the static component
only up to 04:56:00 UT. The blue-shifted emission appeared at
04:56:00 UT and increased instantly, but  up to 04:56:08 UT it is
dominated by the static component. After that the spectrum was
dominated by broadened blue-shifted emission entirely. The observed
(yellow) and synthetic (black) spectra recorded between 04:56:12 UT
and 04:56:18 UT (6 sec accumulation time) as well as blue-, red- and
stationary (green line) components of the synthetic spectra are
shown in Fig. 6 upper-left panel. While blue-shifted synthetic
component is narrower then much more dispersed observed one, the
overall agreement between model and observations seems to be
fulfilled. The blue-shifted emission diminished gradually after
04:54:30 UT. Faint red-shifted emission appeared in the synthetic
spectra between 04:56:24 UT and 04:56:29 UT. In Fig. 6 (right panel)
we compared the observed and synthetic spectra for the period
(between 04:56:30 UT and 04:56:36 UT with 6 sec accumulation time).
Again, the  qualitative agreement between theory and observations is
good. Blue-shifted component vanished entirely after 04:56:46 UT
($\sim 20 - 30$ sec earlier than observed), and BCS CaXIX spectrum
became static again.

\subsection{C2.7 flare on 1992 February 2}

HXT/LO images of the flare show the X-ray emission arriving from a
single loop with semi-length $\sim 9 750$ km, inclined to the local
vertical direction by 35 deg (see Fig. 1, central panel).

The numerical model of the flare was calculated for the period from
11:32:32 UT to 11:35:57 UT (205 sec), covering an impulsive phase
and spreading beyond maximum of the emission recorded with
\emph{GOES}. Stationary component was present in both synthetic and
observe spectra up to 11:33:01 UT only . There is a good compliance
between modeled and observed spectra although the registered signal
was weak. After 11:33:13 UT the blue-shifted component with velocity
v$\sim250$ km/s appeared in the spectra. Initially weak, this
component increased instantly.

The blue-shifted component dominated after 11:33:19 UT the
stationary one. The velocity of blue-shifted component was about
v$\sim 220$ km/s. There is good agreement between model and observe
spectra (see Fig. 6 - left panel, where the color line markings are
the same as in earlier events). The stationary component dominates
from 11:33:31 UT again and mean velocity of the blue-shifted
component decreased to v$= 190$ km/s. The time evolutions and shapes
of synthetic and observed spectra match well. The blue-shifted
emission in the modeled spectrum diminished after 11:33:36 UT
gradually, vanishing entirely at 11:33:43 UT, when the synthetic BCS
CaXIX spectrum became static again. No red-shifted emission was
present during the whole modeled period of the flare's evolution.

\subsection{M2.4 flare on 2000 July 27}

The flare was visible in SXT as a single loop perpendicular to the
solar surface (see Fig. 1, right panel) with semi-length $\sim 8
690$ km.

The numerical model of the flare was calculated for the period from
04:07:56 UT to 04:10:30 UT, covering an impulsive phase and maximum
of the emission recorded with \emph{GOES}.  The observed BCS CaXIX
spectra include blue-shifted and stationary components up to
04:07:46 UT.  The blue-shifted component with velocity v = 180 km/s
(which is two times weak than stationary) can be seen in the
observed spectra. The model agrees well with the observed spectra
although the observed lines are more broadened that the ones in the
model. The blue-shifted component with velocity v$=290$ km/s
dominates in the spectra after 04:08:10 UT. There is a good
compliance between modeled and observed spectra. In this time also a
faint red-shifted emission was present on the synthetic spectra
(Fig. 6 - left panel, the color of the lines are the same as in
earlier events). After 04:08:16 UT the stationary component begins
to dominate again and velocity of blue component decreased (v=120
km/s), until complete disappearance (see Fig. 6 - right panel). A
faint red-shifted emission was still present on the synthetic
spectra. After 04:08:43 UT  the synthetic BCS CaXIX spectrum become
static again.

\section{Discussion and conclusions}

Numerical modeling of the flares allows to study
distributions of the plasma velocities along the flaring loops.
Thus, it is possible to estimate relative contributions of the
emissions of the stationary and moving plasma to the spectra (i.e.
their stationary and shifted components).

Most of papers describing results of the analysis of the BCS
data indicate that the BCS spectra are dominated by stationary component.
Additionally, there is only a very limited number of described
observations of the solar flares having spectra
in which the blue-shifted component dominates the stationary one.
The statistical paper by Gan and Watanabe shows that although moderate
blue asymmetry is very common during the impulsive phase, the number
of events having spectra with great blue-shift is very small. They also found that
for most flares the blue-shift emission appears during early
impulsive phases and is temporally correlated with the line
broadening.

Most 1D numerical models of the chromospheric evaporation in
single-loops show a presence of very fast up-flows, up to several
hundred kilometers per second. This discrepancy between observations
and models could be explained by insufficient sensitivity of the
instruments but most of the theoretical works present models of the
solar flare having low up-flow velocities. The most advanced model
of this kind is a multi-thread model of the flare proposed by Hori
and co-workers and developed by Doschek and Warren,
consistent with the observed BCS spectra. In this paper we
present other explanation of the lack of the observed blue-shifted
spectra.

 We investigated C2.7 \emph{GOES} class
flare observed on 1992 February 2, M2.4 flare observed on 2000 July
27, and M2.7 class flare observed on 1991 December 16. Two of the
analyzed flares (C2.7 and M2.4) were not described previously in
literature; the third one, M2.7 was investigated by
\citet{Culhane94}. The synthetic BSC spectra of the flares were
calculated using real observational parameters of the observed
flares (geometry, physical parameters of the plasma, energy spectrum
of the non-thermal electrons).

For all investigated events we obtained good quantitative
conformability of the synthetic and observed BCS CaXIX spectra. For
the whole impulsive phase of each flare we obtained overall good
agreement of fluxes, shapes and time variations of the observed and
synthetic spectra.

Generally, widths of the observed spectral lines were greater than
widths of the relevant synthetic spectra lines. The observed
blue-shifted component was wider than synthetic one and highly
asymmetric. This suggests wider dispersion of up-flow velocities in
observed events than in our theoretical models. This
non-conformability could be caused by disregarded turbulent motions
of the plasma in our numerical model and by some differences between
calculated and real distributions of the physical parameters of the
plasma along the loops (densities, temperatures, and macroscopic
velocities). Nevertheless, we obtained good conformity between
overall shape, stationary, red- and blue-shifted components of the
relevant calculated and observed spectra.

Despite very high up-flow velocities of the plasma obtained in
numerical simulations of the various flares similarly high radial
 velocities could be observed mostly for the events located in the
central part of the solar disc only, due to a geometrical projection
of the velocity vector onto the direction of sight. Assuming a
semi-circular loop located perpendicularly to the local solar
surface and rooted at the solar longitude of 60 deg the LOS
component of the velocity is two times smaller than the real
velocity of the plasma along the loop. Similarly about 50\% of the
emission emitted as blue-shifted from the loop rooted at the disc
centre is observed as a stationary component for the loop rooted at
the solar longitude of 60 deg. So, taking into account the
geometrical dependences of the LOS velocities of the plasma moving
along the loops inclined toward the solar surface as well as a
distribution of the flare sites over the solar disc we can conclude
that stationary component can be observed for all flares during
their early phases of evolution (i.e., in almost all flares exist
some plasma which don't move along the LOS). On the other side, the
blue-shifted component of the spectrum could be not detected even
for plasma rising along the flaring loop with very high velocity.
Additionally our simulations based on realistic heating rates by
non-thermal electrons indicate that the upper chromosphere plasma is
heated by non-thermal electrons a few seconds before beginning of
noticeable high-velocity bulk motions, and until this time its emits
stationary (not red- or blue-shifted) spectrum only.

Very similar investigations of the impact of the various
viewing angles on the resulting spectral signature of the flare was
made already by \citet{Li}. Li and co-workers used theoretical
models of flaring loops with arbitrary selected plasma parameters
and temporal variations of the local heating. For most cases they
obtained a strong blue-shifted component of the spectrum. Our models
are based on observational data; especially duration of the heating
and delivered energy flux were evaluated from HXT/\emph{Yohkoh}
data. In our models the absolute velocities of the plasma
up-flow were of the order of 450-500 km/s, but taking into account
the geometrical effect of the angle of view on the loops they were
reduced to moderate 200-250 km/s.

Taking into account the geometrical dependences of the
line-of-sight velocities of the plasma moving along the flaring loop
inclined toward the solar surface as well as a distribution of the
investigated flares over the solar disk, we conclude that stationary
component of the spectrum should be observed almost for all flares
during their early phases of evolution. In opposite, the
blue-shifted component of the spectrum could be not detected in flares
having plasma rising along the flaring loop even with high velocity
due to the geometrical dependences only. Our simulations based on
realistic heating rates of plasma by non-thermal electrons indicate
also that the upper chromosphere is heated by non-thermal electrons
a few seconds before beginning of noticeable high-velocity bulk
motions, and before this time plasma emits stationary component of
the spectrum only. After the start of the up-flow motion, the
blue-shifted component dominate temporally the synthetic spectra of
the investigated flares at their early phases.

\begin{acknowledgements}
The authors would like to thank \emph{Yohkoh} team for excellent
solar data and software. They are also grateful to the anonymous
referee for useful comments and suggestions. This work was supported by the Polish
Ministry of Science and Higher Education, grant number N203 022
31/2991.
\end{acknowledgements}


\end{document}